\documentstyle[12pt]{article}
\begin{document}
\title{Stationary Cylindrical Anisotropic Fluid}

\author{F. Debbasch$^1$\thanks{e-mail: debbasch@lra.ens.fr}, L.
Herrera$^2$\thanks{e-mail: laherrera@movistar.net.ve}, \\
P. R. C. T. Pereira$^3$\thanks{e-mail: terra@cbpf.br} and N. O.
Santos$^{1,4,5}$\thanks{e-mail: santos@ccr.jussieu.fr and
N.O.Santos@qmul.ac.uk}\\
\small{$^1$Universit\'e Pierre et Marie Curie - CNRS/FRE 2460,} \\
\small{LERMA/ERGA, Tour 22-12, 4\`eme \'etage, Bo\^{\i}te 142, 4
place
Jussieu,} \\
\small{75252 Paris Cedex 05, France.} \\
\small{$^2$Escuela de F\'{\i}sica, Facultad de Ciencias,} \\
\small{Universidad Central de Venezuela, Caracas, Venezuela.} \\
\small{$^3$Centro Brasileiro de Pesquisas F\'{\i}sicas, 22290-180
Rio de
Janeiro RJ, Brazil.} \\
\small{$^4$School of Mathematical Sciences, Queen Mary,}\\
\small{University of London, London E1 4NS, UK.}\\
\small{$^5$Laborat\'orio Nacional de Computa\c{c}\~ao
Cient\'{\i}fica, 25651-070 Petr\'opolis RJ, Brazil.}} \maketitle

\begin{abstract}
We present the whole set of equations with regularity and matching
conditions required for the description of physically meaningful
stationary cylindrically symmmetric distributions of matter,
smoothly matched to Lewis vacuum spacetime. A specific example is
given. The electric and magnetic parts of the Weyl tensor are
calculated, and  it is shown that purely electric solutions are
necessarily static. Then, it is shown that  no conformally flat
stationary cylindrical fluid exits, satisfying regularity and
matching conditions.
\end{abstract}

\newpage
\section{Introduction}
In a recent paper \cite{Herrera} we have presented a systematic study
of static cylindrically symmetric matter distributions.
To justify such endeavour we mentioned, on the one hand the still ongoing
discussion about the  precise
meaning of the independent parameters of the corresponding vacuum
spacetime, and on the other, the renewed interest in cylindrically
symmetric
sources in relation with different, classical and quantum, aspects of
gravitation
(see \cite{1} and references therein).

It is clear that the extension of the previous study to the stationary
case, is not only justified by the same arguments, but in addition allows
for studying the role of the rotation
of the source.

However, it should be stressed that in spite of the simplifications
introduced by the cylindrical symmetry, a purely analytical handling of the
problem requires further restrictions.
Accordingly, we shall consider here the shear--free (rigid rotation) case.

As in \cite{Herrera} we want to bring out the relationship between
the Weyl tensor and different aspects of the source. This last
question is in turn motivated by the very conspicuous link
existing in the spherically symmetric case between the Weyl
tensor, the inhomogeneity of the energy density and the anisotropy
of pressure \cite{est}. For doing so we have calculated the Weyl
components as well as the components of its electric and magnetic
parts. Unfortunately, no such link can be established in our case.
However it is shown that purely electric solutions are necessarily static.
This last
result, together with results obtained in \cite{Herrera} imply
that  no conformally flat stationary cylindrical fluid exists,
satisfying regularity and matching condition.

The paper is organized as follows: in section 2 we present the
general form of the energy momentum tensor, the line element, the
Einstein equations, the regularity conditions and the active
gravitational mass. The exterior space-time as well as junction
conditions are discussed in section 3. In section 4 a specific
example, which we believe is new, is presented. In section 5 the
electric and magnetic parts of Weyl tensor are presented and the
non existence of conformally flat models satisfying Darmois
conditions is established. Finally, some conclusions are presented
in the last section and the components of Weyl tensor are given in
an appendix.
\newpage

\section{Inside the source}
We shall first describe the kind of source we are dealing with as well as
its spacetime. Then we shall write the corresponding Einstein equations and
discuss the regularity conditions
to assure the correct behaviour of the solutions. Finally we present a very
compact expression for the active gravitational mass of the source.

\subsection{The source}
We consider a stationary cylindrically symmetric anisotropic
non-dissipative fluid bounded by a cylindrical surface $\Sigma$ and with
energy momentum tensor given by
\begin{equation}
T_{\alpha\beta}=(\mu +
P_r)V_{\alpha}V_{\beta}+P_rg_{\alpha\beta}+(P_{\phi}-P_r)K_{\alpha}K_{\beta}
+(P_z-P_r)S_{\alpha}S_{\beta}, \label{1}
\end{equation}
where, $\mu$ is the energy density, $P_r$, $P_z$ and $P_{\phi}$ are the
principal stresses and $V_{\alpha}$, $K_{\alpha}$ and $S_{\alpha}$ are
four-vectors satisfying
\begin{equation}
V^{\alpha}V_{\alpha}=-1, \;\; K^{\alpha}K_{\alpha}=S^{\alpha}S_{\alpha}=1, \;\;
V^{\alpha}K_{\alpha}=V^{\alpha}S_{\alpha}=K^{\alpha}S_{\alpha}=0. \label{2}
\end{equation}

\subsection{The interior spacetime}
We assume for interior metric to $\Sigma$ the general stationary
cylindrically symmetric line element, which can be written
\begin{equation}
ds^2=-fdt^2+2kdtd\phi+e^{\gamma}(dr^2+dz^2)+ld\phi^2, \label{3}
\end{equation}
where $f$, $k$, $\gamma$ and $l$ are all functions of $r$. To represent
cylindrical symmetry, we impose the following ranges on the coordinates
\begin{equation}
-\infty\leq t\leq\infty, \;\; 0\leq r, \;\; -\infty<z<\infty, \;\;
0\leq\phi\leq 2\pi. \label{4}
\end{equation}
We number the coordinates $x^0=t$, $x^1=r$, $x^2=z$ and $x^3=\phi$ and we
choose the fluid being at rest in this coordinate system, hence from
(\ref{2}) and (\ref{3}) we have
\begin{equation}
V_{\alpha}=-\sqrt{f}\delta_{\alpha}^0+\frac{k}{\sqrt{f}}\delta_{\alpha}^3,
\;\; S_{\alpha}=e^{\gamma/2}\delta_{\alpha}^2, \;\;
K_{\alpha}=\left(l+\frac{k^2}{f}\right)^{1/2}\delta_{\alpha}^3.
\label{5}
\end{equation}

It is worth noticing that we are assuming that our coordinates are
co--rotating with the fluid
($V^{\alpha}=\delta^{\alpha}_{0}/\sqrt{f}$), which implies, as it
can be very easily shown, that the fluid is shear--free. Thus we
shall consider a shear--free, expansionless fluid distribution.

\subsection{The Einstein equations}
For the Einstein field equations,
$G_{\alpha\beta}=\kappa T_{\alpha\beta}$, with (\ref{1}), (\ref{3}) and
(\ref{5}) we have the non-null components
\begin{eqnarray}
G_{00}=-\frac{1}{4D^2e^{\gamma}}\left[2fD^2\gamma^{\prime\prime}+4fDD^{\prime\prime}-
2D^2f^{\prime\prime}+2DD^{\prime}f^{\prime} \right. \nonumber \\
\left.-3f(f^{\prime}l^{\prime}+k^{\prime 2})\right]=
\kappa\mu f, \label{6} \\
G_{03}=\frac{1}{4D^2e^{\gamma}}\left[2kD^2\gamma^{\prime\prime}+4kDD^{\prime
\prime}-2D^2k^{\prime\prime}+2DD^{\prime}k^{\prime} \right. \nonumber \\
\left. -3k(f^{\prime}l^{\prime}+k^{\prime 2})\right]=-\kappa\mu k, \label{7} \\
G_{11}=\frac{1}{4D^2}(2DD^{\prime}\gamma^{\prime}+f^{\prime}l^{\prime}
+k^{\prime 2})=\kappa P_re^{\gamma}, \label{8} \\
G_{22}=-\frac{1}{4D^2}(2DD^{\prime}\gamma^{\prime}-4DD^{\prime\prime}
+f^{\prime}l^{\prime}+k^{\prime 2})
=\kappa P_ze^{\gamma}, \label{9} \\
G_{33}=\frac{1}{4D^2e^{\gamma}}\left[2lD^2\gamma^{\prime\prime}+4lDD^{\prime
\prime}-2D^2l^{\prime\prime}+2DD^{\prime}l^{\prime} \right. \nonumber \\
\left. -3l(f^{\prime}l^{\prime}+k^{\prime 2})\right]=
\frac{\kappa}{f}(\mu k^2+P_{\phi}D^2). \label{10}
\end{eqnarray}
where the primes stand for differentiation with respect to $r$ and
\begin{equation}
D^2=fl+k^2. \label{11}
\end{equation}

Thus we have five equations for the eight unknown functions,
namely $f$, $k$, $\gamma$, $l$, $\mu$, $P_r$, $P_z$ and $P_\phi$.
Therefore three equations of state relating the matter variables
or {\it ad hoc} assumptions on the metric functions have to be
imposed in order to integrate (\ref{6})--(\ref{10}) (e.g. see the
example in section 4 below).

The Bianchi identity $T_{r ;\beta}^{\beta}=0$ from
(\ref{1}) reduces to,
\begin{equation}
P^{\prime}_r+\frac{1}{2}(\mu+P_r)\frac{f^{\prime}}{f}+(P_r-P_{\phi})\frac{D^
{\prime}}{D}+\frac{1}{2}(P_r-P_z)\gamma^{\prime}=0.
\label{11a}
\end{equation}
From (\ref{6}) and (\ref{7}) we have
\begin{equation}
kG_{00}+fG_{03}=\frac{D}{2e^{\gamma}}\left(\frac{kf^{\prime}-fk^{\prime}}{D}
\right)^{\prime}=0,
\label{12}
\end{equation}
which allows to be integrated, producing
\begin{equation}
kf^{\prime}-fk^{\prime}=c_1D, \label{13n}
\end{equation}
where $c_1$ is an integration constant.

\subsection{Regularity conditions}
Regularity conditions on the symmetry axis imply \cite{Mac}
\begin{equation}
\lim_{r\to 0}\frac{g^{\alpha\beta}X_{,\alpha} X_{,\beta}}{4X}=1
\label{X}
\end{equation}
with $X=g_{\phi \phi}$.
Then (\ref{X}) yields
\begin{equation}
\lim_{r\to 0}\frac{e^{-\gamma} l^{\prime 2}}{4l}=1,  \label{l}
\end{equation}
The requirement that $g_{\phi \phi} =0$ on the axis, gives
\begin{equation}
l\stackrel{0}{=}0, \label{l2}
\end{equation}
where $\stackrel{0}{=}$ means that the equation is evaluated at
$r=0$.

Since along the axis physically there cannot be singularities, we
impose that spacetime tends near the axis to a flat spacetime,
hence we scale the coordinates so that as $r\rightarrow 0$ we have
\begin{equation}
ds^2=-dt^2+2\omega r^2dtd\phi+dr^2+dz^2+r^2d\phi^2, \label{ds}
\end{equation}
from which it follows that
\begin{equation}
f\stackrel{0}{=}1, \;\; k\stackrel{0}{=}\gamma\stackrel{0}{=}0, \label{13abis}
\end{equation}
implying
\begin{equation}
D\stackrel{0}{=}0, \label{D}
\end{equation}
and from (\ref{l}) and (\ref{ds})
\begin{equation}
l^\prime\stackrel{0}{=}0. \label{l'}
\end{equation}
Finally, from the above and the requirement that Einstein tensor components in
(\ref{6}-\ref{10}) do not diverge, it follows that
\begin{equation}
f^\prime \stackrel{0}{=}k^{\prime}\stackrel{0}{=} k^{\prime
\prime}-k^{\prime}\frac{D^{\prime}}{D}\stackrel{0}{=}0.
\label{der}
\end{equation}

\subsection{The Whittaker mass}
The Whittaker formula \cite{Whittaker} for the active mass per
unit length $m$ of a stationary distribution of perfect fluid with
energy density $\mu$ and principal stresses $P_r$, $P_z$ and
$P_{\phi}$ inside a cylinder of surface $\Sigma$ is
\begin{equation}
m=2\pi \int^{r_{\Sigma}}_0(\mu+P_r+P_z+P_{\phi})\sqrt{-g}dr, \label{14}
\end{equation}
where $g$ is the determinant of the metric. Now substituting (\ref{3}) and
(\ref{6}-\ref{10}) we obtain
\begin{equation}
m=\frac{2\pi}{\kappa}\int^{r_{\Sigma}}_0\left(\frac{lf^{\prime}
+kk^{\prime}}{D}\right)^{\prime}dr=\frac{2\pi}{\kappa}\int^{r_{\Sigma}}_0
\left(\frac{Df^{\prime}-c_1k}{f}\right)^{\prime}dr, \label{15}
\end{equation}
where we have used (\ref{13n}). Integrating (\ref{15}) and using the regularity
conditions on the axis we obtain
\begin{equation}
m\stackrel{\Sigma}{=}\frac{2\pi}{\kappa}\frac{Df^{\prime}-c_1k^{\prime}}{f}.
\label{15a}
\end{equation}

\section{Exterior spacetime and junction conditions}
For the exterior spacetime of the cylindrical surface $\Sigma$,
since the system is stationary, we take the Lewis metric
\cite{Lewis} and consider its Weyl class \cite{vor},
\begin{equation}
ds^2=-Fdt^2+2Kdtd\phi+e^{\Gamma}(dR^2+dz^2)+Ld\phi^2, \label{a26}
\end{equation}
where
\begin{eqnarray}
F=aR^{1-n}-a\delta^2R^{1+n}, \label{a27} \\
K=-(1-ab\delta)\delta R^{1+n}-abR^{1-n}, \label{a28} \\
e^{\Gamma}=R^{(n^2-1)/2}, \label{a28} \\
L=\frac{(1-ab\delta)^2}{a}R^{1+n}-ab^2R^{1-n}, \label{a29}
\end{eqnarray}
with
\begin{equation}
\delta=\frac{c}{an}, \label{a30}
\end{equation}
and $a$, $b$, $c$ and $n$ are real constants. The coordinates $t$,
$z$ and $\phi$ can be taken as in (\ref{3}) and with the same
ranges (\ref{4}). The radial coordinates in (\ref{3}) and
(\ref{a26}), $r$ and $R$, are not necessarily continuous on
$\Sigma$ in order to preserve the range of $\phi$ in (\ref{4}).
Also, since we are considering an expansionless cylinder and the
fluid is at rest in our coordinate system, the equation of the
boundary surface should read $r\stackrel{\Sigma}{=}$ constant and
$R\stackrel{\Sigma}{=}$ constant, from inside and outside,
respectively.

In accordance with Darmois junction conditions \cite{Darmois} we
obtain that the metric coefficients (\ref{3}) and ({\ref{a26}) and
its derivatives must be continuous across $\Sigma$,
\begin{eqnarray}
f\stackrel{\Sigma}{=}F, \;\; k\stackrel{\Sigma}{=}K, \;\;
\gamma\stackrel{\Sigma}{=}\Gamma, \;\; l\stackrel{\Sigma}{=}L, \label{b30} \\
\frac{f^{\prime}}{f}\stackrel{\Sigma}{=}\frac{1}{R}+n\frac{\delta^2R^n+R^{-n
}}{\delta^2R^{1+n}-R^{1-n}},
\label{b31}\\
\frac{k^{\prime}}{k}\stackrel{\Sigma}{=}\frac{1}{R}+n\frac{(1-ab\delta)\delta
R^n-abR^{-n}}{(1-ab\delta)\delta R^{1+n}+abR^{1-n}}, \label{b32} \\
\gamma^{\prime}\stackrel{\Sigma}{=}\frac{n^2-1}{2R}, \label{b33} \\
\frac{l^{\prime}}{l}\stackrel{\Sigma}{=}\frac{1}{R}+n\frac{(1-ab\delta)^2R^n
+a^2b^2R^{-n}} {(1-ab\delta)^2R^{1+n}-a^2b^2R^{1-n}}. \label{b34}
\end{eqnarray}
hence we have from (\ref{8}) $P_r\stackrel{\Sigma}{=}0$, as
expected, and for (\ref{13n}) and (\ref{15a})
\begin{eqnarray}
c_1=2c, \label{a31} \\
m=\sigma+\frac{bc}{2}, \label{a32}
\end{eqnarray}
where $\sigma$ is the Newtonian mass per unit length, as measured
outside the source, given by
\begin{equation}
\sigma=\frac{1-n}{4}. \label{a33}
\end{equation}
The two parameters $b$ and $c$ are responsible for the
stationarity of the spacetime, accordingly the second term in
(\ref{a32}) may be interpreted as the ``rotational'' contribution
to the active gravitational mass of the source.

For a
cylindrically symmetric rigidly rotating dust, van Stockum
solution \cite{Stockum}, one obtains
\begin{equation}
bc=-\frac{(1-n)^2}{4}, \label{aa33}
\end{equation}
or with (\ref{a32}-\ref{aa33})
\begin{equation}
m=\sigma(1-2 \sigma), \label{mnu}
\end{equation}
which could be interpreted as if, in this case, rotation
diminishes the active gravitational mass. At first sight this
result seems counterintuitive, as we should expect the energy of
the rotational motion to increase the total energy of the source.

 However, the
situation is not so simple. First of all observe that the van
Stockum solution has no  static limit (see next section). Indeed,
reducing its rotation to zero amounts to delete the source (the
spacetime becomes flat). Therefore it is not clear that in this
case, $m$ as given by (\ref{a32}) might be interpreted as
describing the static plus the ``rotational'' contributions  to
the active gravitational mass of the source, since no static
contribution exists for this case.

In the same line of arguments it is worth noticing that circular
geodesics of test particles become null when $\sigma =1/4$, for
both, the static and stationary cylinder \cite{HS}, but this
corresponds to $m=1/4$ in the first case and $m=1/8$ in the van
Stockum case.

On the other hand, we know of at least one example where the angular
momentum of the source diminishes the effective active gravitational mass
of the source: the Kerr metric.
Indeed, in this later case the equation for radial geodesic on the symmetry
axis reads:
\begin{equation}
\ddot r=-M\frac{r^2-a^2}{(r^2+a^2)^2}
\label{kerr}
\end{equation}
where the dot stands for differentiation  with respect to the
proper time, $r$ is the radial Boyer-Lindquist coordinate and $a$
and $M$ are the two parameters of the Kerr metric. It is evident
from (\ref{kerr}) that the angular momentum of the source $a$
diminishes the attraction force on the test particle, as compared
with the static case.

All this having been said,  we do not know at this point whether or not
some kind of mechanism similar to (or  different from)  the one existing in
the Kerr spacetime is  acting in the
van Stockum case, and more important we do not know what the physics
underlying such mechanism, might be.

\section{An interior solution}
In order to illustrate the use of all the equations above, let us present a
simple model which we believe is new. As we shall see, even in a very
restrictive situation, as the one presented below, a purely analytical
treatment of the problem  is not suitable,
and one should appeal to numerical methods, for a complete description of
the solution. Nevertheless, some information about the properties of the
model may be obtained from the
analytical approach.

Thus, let us assume for simplicity the  equations of state
\begin {equation}
P_z=\alpha\mu, \;\; P_r=P_{\phi}=0, \label{n1}
\end{equation}
with $\alpha =$ constant. When $\alpha=0$ (\ref{n1}) reduces to
the van Stockum solution \cite{Stockum} for a cylindrically
symmetric rotating dust which is given by
\begin{eqnarray}
f=1, \;\; k=\omega r^2, \;\; \gamma=-\omega^2r^2, \;\;
l=r^2(1-\omega^2r^2),
\label{n1a} \\
\kappa\mu=4\omega^2e^{\omega^2r^2}, \label{n1b}
\end{eqnarray}
where $\omega$ is a constant.

Substituting (\ref{n1}) into (\ref{11a}) we have for $\mu\neq 0$
\begin{equation}
\alpha\gamma^{\prime}=\frac{f^{\prime}}{f}, \label{n2}
\end{equation}
and integrating we obtain
\begin{equation}
e^{\alpha\gamma}=c_2f, \label{n2a}
\end{equation}
where $c_2$ is an integration constant. Considering the regularity
conditions (\ref{13abis}) we can rewrite (\ref{n2a}) as
\begin{equation}
e^{\alpha\gamma}=f. \label{n2b}
\end{equation}
The field equations (\ref{6}-\ref{10}) with (\ref{n1}) reduce to
\begin{eqnarray}
2DD^{\prime}\gamma^{\prime}+f^{\prime}l^{\prime}+k^{\prime 2}=0,
\label{n3} \\
4(1+\alpha)fDD^{\prime\prime}+\alpha\left[2fD^2\gamma^{\prime\prime}-2D^2f^{
\prime\prime}+2DD^{\prime}f^{\prime}-3f(f^{\prime}l^{\prime}
+k^{\prime 2})\right]=0. \label{n4}
\end{eqnarray}
Substituting (\ref{13n}), (\ref{a31}) and (\ref{n2}) into
(\ref{n3}) and considering (\ref{11}) we have
\begin{equation}
2(1+\alpha)ff^{\prime}D^{\prime}-\alpha D(f^{\prime 2}-4c^2)=0;
\label{n6}
\end{equation}
while substituting (\ref{n2}) and (\ref{n3}) into (\ref{n4}) we
obtain
\begin{equation}
2(1+\alpha)f^2D^{\prime\prime}+(1-\alpha)Dff^{\prime\prime}-Df^{\prime
2}+(3+\alpha)ff^{\prime}D^{\prime}=0. \label{n7}
\end{equation}
Now from (\ref{n6}) and (\ref{n7}) we have
\begin{equation}
(f^{\prime 2}+4\alpha
c^2)\left[2(1+\alpha)ff^{\prime\prime}-(2+\alpha)f^{\prime
2}+4\alpha c^2\right]=0. \label{n8}
\end{equation}
If $\alpha=0$ then (\ref{n8}) reduces to $f^{\prime}=0$, giving
 the van Stockum's solution (\ref{n1a}). However
if $\alpha\neq 0$ and since
\begin{equation}
f^{\prime 2}+4\alpha c^2\neq 0, \label{n8a}
\end{equation}
considering the transformation
\begin{equation}
f^{\prime 2}=x\neq 0, \label{n9}
\end{equation}
(\ref{n8}) reduces to
\begin{equation}
\frac{f^{\prime}}{f}=\frac{(1+\alpha)x^{\prime}}{(2+\alpha)x-4\alpha
c^2}. \label{n10}
\end{equation}
Integrating (\ref{n10}) we obtain
\begin{equation}
f=c_3\left[(2+\alpha)x-4\alpha c^2\right]^{(1+\alpha)/(2+\alpha)},
\label{n11}
\end{equation}
where $c_3$ is an integration constant. From the regularity
conditions (\ref{13abis}) and (\ref{der}) we can rewrite
(\ref{n11}),
\begin{equation}
f=\left(1-\frac{2+\alpha}{4\alpha
c^2}\;x\right)^{(1+\alpha)/(2+\alpha)},\label{n12}
\end{equation}
or with the aid of (\ref{n2}) it becomes
\begin{equation}
f=\left[1-\frac{(2+\alpha)\alpha}{4c^2}f^2\gamma^{\prime
2}\right]^{(1+\alpha)/(2+\alpha)}. \label{nn12}
\end{equation}
From (\ref{nn12}) we see that when $\alpha=0$ it reduces to the
van Stockum solution (\ref{n1a}).

Feeding back (\ref{n9}) into  (\ref{n12}) we could in principle integrate
to obtain $f=f(r)$, unfortunately though, the integral can only be
expressed in terms of hypergeometric functions, therefore further
specification of the solution is only possible numerically.

However, some general properties of the solution may be found without
performing the integration of (\ref{n12}).

Thus, the matching of (\ref{n2}) or (\ref{n6}) using
(\ref{b30}-\ref{b34}) produces
\begin{equation}
\alpha\stackrel{\Sigma}{=}\frac{2}{n^2-1}\left(1+n\frac{\delta^2R^n+R^{-n}}{
\delta^2R^n-R^{-n}}\right),
\label{nn1}
\end{equation}
showing that $\alpha$ can have positive as well as negative values
for different parameters that describe the source.

Substituting (\ref{n2}), (\ref{n3}-\ref{n6}) and (\ref{n12}) into
(\ref{6}) we obtain
\begin{equation}
\kappa\mu e^{\gamma}=\left[\frac{2c}{(1+\alpha)f}\right]^2.
\label{n13}
\end{equation}
At the axis $r=0$ (\ref{n13}) becomes
\begin{equation}
\kappa\mu\stackrel{0}{=}\left(\frac{2c}{1+\alpha}\right)^2.
\label{nn13}
\end{equation}
If we fix the stationary parameter $c$ we observe from
(\ref{nn13}) that if $\alpha>0$, or $P_z>0$, the density $\mu$
diminishes as compared to the van Stockum dust solution; however
if $\alpha<0$, or $P_z<0$, then $\mu$ increases as compared to the
same solution.

The model cannot be specified further until (\ref{n12}) is integrated and
the full set of boundary conditions are used. However, as mentioned before
this requires the introduction of
numerical procedures, which is out of the scope of the paper.

\section{The electric and the magnetic parts of  Weyl tensor}
The electric and magnetic parts of Weyl tensor, $E_{\alpha \beta}$
and $H_{\alpha\beta}$, respectively, are formed from the Weyl
tensor $C_{\alpha \beta \gamma \delta}$ and its dual $\tilde
C_{\alpha \beta \gamma \delta}$ by contraction with the four
velocity vector $V^{\alpha}$, thus
\begin{equation}
E_{\alpha \beta}=C_{\alpha \gamma \beta \delta}V^{\gamma}V^{\delta}
\label{electric}
\end{equation}
\begin{equation}
H_{\alpha \beta}=\tilde C_{\alpha \gamma \beta
\delta}V^{\gamma}V^{\delta}= \frac{1}{2}\epsilon_{\alpha \gamma
\epsilon \delta} C^{\epsilon \delta}_{\quad \beta \rho} V^{\gamma}
V^{\rho}, \;\; \epsilon_{\alpha \beta \gamma \delta} \equiv
\sqrt{-g} \;\;\eta_{\alpha \beta \gamma \delta} \label{magnetic}
\end{equation}
where $\eta_{\alpha\beta\gamma\delta}= +1$ or $-1$ for $\alpha,
\beta, \gamma, \delta$ in even or odd order, respectively, and $0$
otherwise, and with
\begin{eqnarray}
V^{\alpha}=\frac{\delta^{\alpha}_{0}}{\sqrt{f}}, \label{v} \\
\sqrt{-g}=De^{\gamma}. \label{g}
\end{eqnarray}
Then it follows that there are only three non--vanishing electric
components, namely
\begin{eqnarray}
E_{11}=\frac{C_{0101}}{f} \label{e1}, \\
E_{22}=\frac{C_{0202}}{f} \label{e2}, \\
E_{33}=\frac{C_{0303}}{f} \label{e3},
\end{eqnarray}
however they are not independent, since, by virtue of (\ref{23})  they
satisfy the relation
\begin{equation}
E_{11}+E_{22}=-\frac{fe^\gamma}{D^2}E_{33}. \label{e4}
\end{equation}
Thus, there are two independent electric components of the Weyl tensor.

On the other hand it can be easily checked that there is only one
independent magnetic component, namely
\begin{equation}
H_{12}=\frac{1}{D}\left(C_{0223}-\frac{k}{f}C_{0202}\right).
\label{m1}
\end{equation}
Accordingly, if we demand our spacetime to be purely electric
($H_{\alpha \beta}=0$), then using (\ref{13n}), (\ref{m1}),
(\ref{17}) and (\ref{20}) we obtain
\begin{equation}
\gamma^{\prime}(kf^{\prime}-fk^{\prime})=0, \label{m2}
\end{equation}
implying that either $\gamma^{\prime}=0$ or $kf^{\prime}-fk^{\prime}=0$.
In the first case it follows from the regularity conditions that
$\gamma=0$. Then matching condition  (\ref{b33}) implies $n=1$, however
this last condition
implies because of (\ref{a33}) that  $\sigma=0$, which produces a Minkowski
spacetime outside the cylinder.

On the other hand   $kf^{\prime}-fk^{\prime}=0$ implies, because
of regularity conditions, that $k=0$.

Thus purely electric cylindrically symmetric spacetimes are necessarily
static. This is in agreement with the result by Glass \cite{glass}
indicating that a necessary and sufficient
condition for a shear--free perfect fluid to be irrotational is that the
Weyl tensor be purely electric.  However, our result does not require the
pressure to be isotropic.

Considering the field equations (\ref{6}), (\ref{7}), (\ref{9})
and (\ref{10}) we can rewrite (\ref{e1}), (\ref{e2}) and
(\ref{m1}) as
\begin{eqnarray}
E_{11}=\frac{1}{12}\left[\kappa e^{\gamma}(7\mu +5P_z+5P_{\phi})
+\left(4\frac{D^{\prime}}{D}-3\frac{f^{\prime}}{f}\right)\gamma^{\prime}-\frac{f^{\prime}l^{\prime}
+k^{\prime2}}{D^2}\right],
\label{e11} \\
E_{22}=\frac{1}{12}\left[\kappa e^{\gamma}(\mu+5P_z-P_{\phi})
-\left(2\frac{D^{\prime}}{D}-3\frac{f^{\prime}}{f}\right)\gamma^{\prime}
-\frac{f^{\prime}l^{\prime}+k^{\prime 2}}{D^2}\right],
\label{e22} \\
H_{12}=\frac{1}{4D}\left(k^{\prime}-\frac{f^{\prime}}{f}k\right)\gamma^{\prime},
\label{h12}
\end{eqnarray}
and with (\ref{13n}) we have
\begin{equation}
H_{12}=-\frac{c_1}{4}\frac{\gamma^{\prime}}{f}, \label{1h12}
\end{equation}
or using (\ref{a31})
\begin{equation}
H_{12}=-\frac{c}{2}\frac{\gamma^{\prime}}{f}. \label{1h13}
\end{equation}
Since the parameter $c$ can be shown to be proportional to the
vorticity of the source \cite{vor}, this last equation reinforces
further the belief that rotation is one of the sources of the
magnetic part of Weyl tensor \cite{BW}. However, it should be
noticed that spacetimes are known which have vorticity but zero
$H_{\alpha \beta}$ \cite{Collins}. The most conspicuous example of
this kind of spacetime being the G\"{o}del solution
\cite{maartens}.

\section{Conclusions}
We have deployed the equations describing the stationary (shear--free) cylinder,
as well as the regularity and matching conditions and the active
gravitational mass. Then a simple model has been presented to
illustrate the way in which models can be found.

Next, we have calculated the Weyl tensor components, as well as its
electric and magnetic parts. It has been established that purely electric
solutions are necessarily static. This
result, together with a previously known result \cite{Herrera} about the
nonexistence of  (static) conformally flat
solution  which satisfies regularity conditions and matches smoothly to
Levi-Civita spacetime on the boundary surface, implies the nonexistence of
interior (stationary) conformally flat
solution  which satisfies regularity conditions and matches
smoothly to
Lewis spacetime on the boundary surface.

Finally,  a link between
the magnetic part of the Weyl tensor and a parameter related to the
vorticity of the source has been found.

\section*{Acknowledgments}
NOS gratefully acknowledges financial assistance from EPSRC United
Kingdom and CNPq Brazil.

\section*{Appendix}
The spacetime (\ref{3}) has the following non-null components of
the Weyl tensor $C_{\alpha\beta\gamma\delta}$,
\begin{eqnarray}
C_{0101}=-\frac{1}{12}\left[f\gamma^{\prime\prime}-3\left(f\frac{D^{\prime}}{D}-
f^{\prime}\right)\gamma^{\prime}+2f\frac{D^{\prime\prime}}{D}
\right.
\nonumber \\
\left. -3f^{\prime\prime}+ 3\frac{D^{\prime}}{D}f^{\prime}-
2\frac{f}{D^2}(f^{\prime}l^{\prime}+k^{\prime 2})\right], \label{16} \\
C_{0202}=-\frac{1}{12}\left[f\gamma^{\prime\prime}+3\left(f\frac{D^{\prime}}
{D}-f^{\prime}\right)\gamma^{\prime}
-4f\frac{D^{\prime\prime}}{D} \right. \nonumber \\
\left. +3f^{\prime\prime}-3\frac{D^{\prime}}{D}f^{\prime}+
4\frac{f}{D^2}(f^{\prime}l^{\prime}+k^{\prime 2})\right], \label{17} \\
C_{0303}=-D^2e^{-2\gamma}C_{1212}=
\frac{D^2e^{-\gamma}}{6}\left[\gamma^{\prime\prime}-\frac{D^{\prime\prime}}{D}
+\frac{1}{D^2}(f^{\prime}l^{\prime}+k^{\prime 2})\right], \label{18} \\
C_{0113}=-\frac{1}{12}\left[k\gamma^{\prime\prime}-3\left(k\frac{D^{\prime}}
{D}-k^{\prime}\right)\gamma^{\prime}+2k\frac{D^{\prime\prime}}{D}
\right.
\nonumber \\
\left. -3k^{\prime\prime}+
3\frac{D^{\prime}}{D}k^{\prime}-2\frac{k}{D^2}(f^{\prime}l^{\prime}
+k^{\prime 2})\right], \label{19}  \\
C_{0223}=-\frac{1}{12}\left[k\gamma^{\prime\prime}+
3\left(k\frac{D^{\prime}}{D}-k^{\prime}\right)\gamma^{\prime}
-4k\frac{D^{\prime\prime}}{D} \right. \nonumber \\
\left. +3k^{\prime\prime}-
3\frac{D^{\prime}}{D}k^{\prime}+4\frac{k}{D^2}(f^{\prime}l^{\prime}
+k^{\prime 2})\right], \label{20} \\
C_{1313}=\frac{1}{12}\left[l\gamma^{\prime\prime}-3\left(l\frac{D^{\prime}}{D}-
l^{\prime}\right)\gamma^{\prime}+2l\frac{D^{\prime\prime}}{D}\right.
\nonumber \\
\left. -3l^{\prime\prime}+3\frac{D^{\prime}}{D}l^{\prime}-
2\frac{l}{D^2}(f^{\prime}l^{\prime}+k^{\prime 2})\right], \label{21} \\
C_{2323}=\frac{1}{12}\left[l\gamma^{\prime\prime}+3\left(l\frac{D^{\prime}}{D}-
l^{\prime}\right)\gamma^{\prime}-4l\frac{D^{\prime\prime}}{D} \right.
\nonumber \\
\left.
+3l^{\prime\prime}-3\frac{D^{\prime}}{D}l^{\prime}+4\frac{l}{D^2}(f^{\prime}
l^{\prime}+k^{\prime 2}) \right], \label{22}
\end{eqnarray}
and they satisfy the relations,
\begin{eqnarray}
C_{0101}+C_{0202}=-\frac{fe^{\gamma}}{D^2}C_{0303}, \label{23} \\
C_{0113}+C_{0223}=-\frac{ke^{\gamma}}{D^2}C_{0303}, \label{24} \\
C_{1313}+C_{2323}=\frac{le^{\gamma}}{D^2}C_{0303}, \label{25} \\
lC_{0101}+fC_{2323}=k(C_{0223}-C_{0113}), \label{b26}
\end{eqnarray}
hence we have  three independent Weyl tensor components, which is
one more than the number in the static case \cite{Herrera}.

\end{document}